\documentstyle[epsfig,12pt]{article}
\newcommand{\be}{\begin{equation}} 
\newcommand{\ee}{\end{equation}}
\newcommand{\bea}{\begin{eqnarray}}
\newcommand{\eea}{\end{eqnarray}}
\newcommand{\bean}{\begin{eqnarray*}}
\newcommand{\eean}{\end{eqnarray*}}
\newcommand{\ba}{\begin{array}}
\newcommand{\ea}{\end{array}}

\newcommand{\norsl}{\normalsize\sl}
\newcommand{\norsc}{\normalsize\sc}
\newcommand{\prd}{Phys.\ Rev.\ D}
\newcommand{\prl}{Phys.\ Rev.\ Lett.}
\newcommand{\plb}{Phys.\ Lett.\ {\bf B}}
\newcommand{\npb}{Nucl.\ Phys.\ {\bf B}}

\def\jp{J/\psi}

\def\ra{\rightarrow}
\begin{document}

\begin{titlepage}

\title{$\jp + c + \bar{c}$ Photoproduction in $e^+\ e^-$ Scattering}

\author{
\norsc  Cong-Feng Qiao\\
\norsl  CCAST(World Lab.), P.O. Box 8730, 
        Beijing 100080, CHINA\\
\norsl  and\\
\norsl  Dept. of Physics, Graduate School of the Chinese 
        Academy of Sciences\\
\norsl  YuQuan Road 19A, Beijing 100039, CHINA\\
\\
\norsc  Jian-Xiong Wang\\
\norsl  Institute of High Energy Physics\\
\norsl  Academia Sinica, Beijing 100039, CHINA}

\date{}
\maketitle

\begin{abstract}
{\normalsize
\noindent
We investigate the $\jp$ + c + $\bar{c}$ photoproduction 
in $e^+\ e^-$ collision at the LEP II energy. 
The physical motivations for this study are: 
1) such process was not considered in previous 
investigations of $\jp$ photoproduction in $e^+\ e^-$ interaction,
and we show in this work that it is worthwhile to do so in order to 
make sound predictions for experimental comparison; 
2) from recent Belle experiment results, the process with same final 
states at the $B$ factory has a theoretically yet unexplainable large 
fraction; hence it is interesting to see what may happen at other 
colliders; 
3) The process can be measured with high accuracy at the planed Linear 
Colliders(LCs);
4) it is necessary to take this process into consideration in the 
aim of elucidating the quarkonium production mechanism, especially 
in testing the universality of NRQCD nonperturbative matrix elements. 
We find that the concerned process is really important at the LEP
experiment energy; within the theoretical uncertainties, it is of 
similar magnitude as the other color-singlet processes when
transverse momentum $p_T > 1$ GeV. Nevertheless, to explain the
recent DELPHI experimental result color-octet mechanism is still 
necessary, but with a shrunk contribution from previous analysis. 
And, it is found that $\jp$ + c + $\bar{c}$ photoproduction process 
can not be mimicked by the simple fragmentation scheme.
}
\end{abstract}

\begin{picture}(5,2)(30,-730)
\put(360,-100){August \ 2003}
\put(360,-115){hep-ph/0308244}
\end{picture}
 
\thispagestyle{empty}
\end{titlepage}
\setcounter{page}{1}
\baselineskip 18pt 
\section{Introduction}
Quarkonium physics is still an interesting research topic
while the first quarkonium state, $\jp$, has been discovered
for about thirty years. Due to its approximately non-relativistic
nature, the description of the heavy quark and anti-quark system
is one of the simplest applications of QCD. The high precise
experimental results of quarkonium leptonic decays render heavy
quarkonium to play a crucial role in investigating various
phenomena, such as measuring the parton distribution in
hadron-hadron collision, detecting the Quark-Gluon-Plasma
signal and even new physics, etc.
On the other hand, the interplay of perturbative and non-perturbative
Quantum Chromodynamics(QCD) happens in the quarkonium production and
decays, which can therefore stand as probes in investigating the 
non-perturbative nature of QCD.

Quarkonium physics has experienced dramatic advances in recent years, 
among them the current focus in the field has been on the colour-octet  
mechanism(COM) \cite{fleming} which was triggered by the high-$p_T$
$J/\psi$ surplus production discovered by the CDF collaboration at the
Tevatron in 1992 \cite{cdf1,cdf2,cdf3}. The colour-octet scenario was
proposed based on a novel effective theory, the non-relativistic
QCD(NRQCD) \cite{nrqcd}. Having achieved the first-step success in
explaining the CDF data, COM imposes as well a strong impact onto
almost every aspects of quarkonium physics, and various efforts have
been made to confirm this mechanism, or to fix the magnitudes of the
universal NRQCD matrix elements. Although the theoretical framework
seems to show qualitative agreements with experimental data, there
are certain difficulties in the quantitative estimate of the
colour-octet contribution \cite{qcf:03}, in particular,
in $\jp$ and $\psi'$ photoproduction at HERA 
\cite{cacciari:yr96,Amundson:yr97,ko:yr96,Kniehl:yr97,kramer:yr96},
$\jp(\psi')$ polarization in large transverse momentum production at
the Fermilab Tevatron \cite{beneke:96yr,braaten:99yr,leibovich:97yr}, 
and more recently in B-factories.

It is widely expected that the B factories would provide clearer 
information of the quarkonium production.
The B-factory experiments recently reported their measurements on the
prompt charmonium production at $e^+e^-$ colliders at $\sqrt{s}=10.6$
GeV \cite{BABAR1,BELLE1,BELLE2}. 
To one's surprise, both their inclusive and exclusive 
measurements have big discrepancies with theoretical calculations 
\cite{p.cho:96,s.baek:98,f.yuan:97,e.braaten:03,k.y.liu:03,k.hagiwara:03}.
Among the puzzling features of the B-factory data, in particular, 
the total cross section of the exclusive 
$e^+ + e^-\rightarrow J/\psi+ \eta_c$ process is found to be
about an order larger than theoretical predictions 
\cite{e.braaten:03,k.y.liu:03,k.hagiwara:03}. 
That is \cite{BELLE2}: 
\begin{equation}
\sigma (e^+ + e^-\rightarrow J/\psi+ \eta_c)\times {\mathcal{B}}
(\eta_c\to \ge 4\mbox{charged})
=(0.033^{+0.007}_{-0.006}\pm 0.009) \mbox{pb}\; , \label{eq:1}
\end{equation}
As well, the Belle collaboration \cite{BELLE2} found a large cross 
section for $J/\psi$ inclusive production along with an open charm
pair, the same final states process as what we are going to discuss,
\begin{equation}
\frac{\sigma (e^+\; e^-\rightarrow J/\psi+\; c\; \bar{c})}
{\sigma (e^+\; e^-\rightarrow J/\psi+\; X)}  = 
0.59^{+0.15}_{-0.13}\pm 0.12\; ,
\end{equation}
which is far more than theoretical expectations 
\cite{p.cho:96,s.baek:98,f.yuan:97}.
The new B-factory data, in
some sense, pose a new "crisis" in the study of qaurkonium physics.  
Therefore, to reveal the problems lying behind the prevailing
quarkonium production mechanisms(models) is currently an urgent task, 
and possibly has a long way to go. Nevertheless, the BELLE "puzzle" 
does not really mean the failure of QCD based quarkonium production
mechanisms, like NRQCD and CS model. The "cirsis" may stem from
the unexplored higher order contributions, for instance, and other
yet unknown reasons within the framework of NRQCD.

The final establishment of NRQCD factorization as the correct theory 
of quarkonium production and decays still needs more tests. The 
universality of NRQCD matrix elements is one of the critical points 
to be verified. People have tried many ways to discover the universality 
of COM at different colliders, as far there is still no decisive result 
either approval or disapproval of it.

Quarkonium photoproduction in $e^+\ e^-$ collisions were investigated by 
several groups \cite{ma:yr98}. And, in very recently, based on leading 
order perturbative QCD analyses, Klasen {\it et al.} \cite{klasen:yr02} 
find that the new DELPHI 
\cite{delphi:yr01} data evidently favor the NRQCD formalism for
$\jp$ production, but rather the conventional colour-singlet(CS) model 
\cite{j.h.kuhn:79},
which is quite encouraging. Considering the data accumulated at 
all LEP four detectors at CERN may still tell us more about the
quarkonium production in the future, we realize it is meaningful to
investigate the quarkonium photoproduction in more details. And, we
find although superficially the $\gamma + \gamma \rightarrow
\jp + c+ \bar{c}$ process stands as a sub-leading order process(in
the sense of strong coupling in comparison with the resolved photon
processes), its contribution to quarkonium production does not
really necessary to be minor to other processes within the CS
prescription. In the direct photon production, the concerned
process here is obviously the leading-contribution process, since in 
experiment the $J/\psi + \gamma$ final state process is suppressed.
In case of resolved photon production, reference \cite{klasen:yr02} 
finds out that single resolved photon processes give the dominant
contribution. The resolved processes in general are suppressed by the
parton distribution probability, but may get compensation from the
order of coupling constant(s). To find whether direct or resolved
proceses dominate in the $J/\psi$ photoproduction at LEP, one needs
to do a concrete calculation.

The reminder of the paper is organized as follows. In Sec. 2, we
will give a description of our calculation procedure. In Sec. 3, our
numerical results are presented, where the theoretical calculation of
$J/\psi + c + \bar{c}$ photoproduction at $e^+\ e^-$ collider is
confronted with recent experimental result at LEP. Finally, we give
our summary and conclusions.

\section{Physics Motivation and Formalism}

As explained in the introduction, we are going to supplement the
process $\gamma + \gamma \ra J/\psi + c\;+\; \bar{c}$ in $e^+\ e^-$
scattering to the analyses of $\jp$ inclusive production at LEP II.
Here, the colliding photons can participate in the hard interaction
either directly, or resolvedly through their hadronic components. 
As realized, both direct and resolved processes can be in the same 
order within our interested energy distribution scope 
\cite{klasen:yr02}. In this sense, since our concerned process is 
perturbatively at sub-leading order relative to the resolved processes 
considered in \cite{klasen:yr02}, it looks to be negligible at first 
glance. However, the complexity of this process makes the order analyses 
not transparent. And further, there is no obvious reason to disregard 
this process while taking color-octet processes into consideration. 
Therefore, to make an overall estimation of $\jp$ photoproduction 
at LEP and draw conclusions without considering the new proposed 
process may run some risk, as shown in next section. Of course, in 
case of having enough events, as inspired by B factory experiments, 
one may expect that the concerned process can be well distinguished 
from other processes. In addition, since the $J/\psi + c\;+\; \bar{c}$ 
final states with a relatively large invariant mass, it is easy to 
imagine that the resolved-photon contribution for this process will 
be less important than the direct one, which is confirmed by our 
numerical evaluation. We find in explicit calculation that the 
resolved-photon one are really negligibly small.

On the other hand, it is easy to attribute the $\gamma + \gamma  
\ra J/\psi + c\;+\; \bar{c}$ 
process to a simple fragmentation representation approximately, where
charmonium is produced via the charm quark fragmentation. It is 
worthwhile to mention that the situation here is different from and 
more complicated than quarkonium production in, e.g., $Z^0$ decays,
where the fragmentation mechanism works pretty well and the calculation
can be greatly simplified by taking the fragmentation limit. Here, some
"nonfragmenting" graphs are not negligible. Explicit numerical result 
given in next section support this argument.

Generally speaking, in photon-photon collision the interacting photons 
can either originate from the bremsstrahlung of high energy 
electron-position collision, beamstralung, or, theoretically, 
be obtained by the Compton back-scattering of 
laser light off the linac electron beams and realize the photon-photon 
collision at a linear collider with approximately the same luminosity 
as that of the $e^+$ $e^-$ beams. Here, in this work we will refrain 
our study on the first case and confront our result with the experimental 
data analyzed recently by DELPHI Collaboration of LEP II experiment 
at CERN. 

The source of photons from electron-position bremsstrahlung can be well
formulated in Weiz\"acker-Williams approximation(WWa) 
\cite{frixione:yr93}:
\bea
\label{eq:wwa}
f_{\gamma/e}(x) = \frac{\alpha_{em}}{2 \pi} \left[\frac{1 + (1 - x)^2}
{x} {\rm log}(Q^2_{max}/Q^2_{min}) + 2 m_e^2\; x \left(\frac{1}{Q^2_{max}}
- 
\frac{1}{Q^2_{min}}\right)\right]\ ,
\eea
where $Q^2_{min} = m_e^2 x^2/(1 - x)^2$ and $Q^2_{max} = (E\ \theta)^2
(1 - x) + Q^2_{min}$ with $x = E_\gamma/E_e$, $\theta$ the experimental
angular cut in order to ensure the photon to be real, and $E = E_e =
\sqrt{s}/2$.

\begin{figure}[tbh]
\begin{center}
\epsfig{file=diag.ps,
bbllx=195pt,bblly=560pt,bburx=410pt, 
bbury=780pt,width=6cm,height=7.5cm,clip=0}
\end{center}
\caption[bt]{Half of the Feynman diagrams of discussed $\jp$ producing 
subprocess, $\gamma + \gamma \rightarrow \jp + c+ \bar{c}$. The missing 
part of diagrams are the charge conjugation of the shown ones, and can be
simply obtained by flipping the fermion flow directions.}
\label{graph2}
\end{figure}    

Our concerned process involves with twenty Feynman diagrams. Ten are 
shown in Figure 1 and the left ten are just the charge conjugation of 
them. It is evident that in this process the quarkonium can be formed
in CS configuration, which may be formulated coincidentally 
in both CS model(CSM) and NRQCD description at the leading order.

The calculation of prompt $\jp$ producing rate will be carried out by 
standard procedure with normalization of spin project operators for the 
quarkonium production taken as: 
\begin{equation}
\label{proj}
{\cal P}_{S,S_z}(P; q) = \sum_{s_1, s_2} v(\frac{P}{2} -q; s_2) \bar{u}
(\frac{P}{2} + q; s_1) <\frac{1}{2}, s_1; \frac{1}{2}, s_2|S, S_z>\ ,
\end{equation}
where $P$ and $S$, $S_z$ are respectively the quarkonium four-momentum, 
its spin and the $z$ component of the spin; $q$ is the relative 
momentum of the heavy quarks; and $s_1$, $s_2$ 
represent their spins. In  the  non-relativistic limit, for S-wave states,
to leading order the covariant forms of the projection operators are very 
simple: 
\begin{eqnarray}
{\cal P}_{0,0}(P; 0) = \frac{1}{2\sqrt{2}}\gamma_5 
(\not\!{P} + M)\ , 
\label{proj0} \\
{\cal P}_{1,S_z}(P; 0) = \frac{1}{2\sqrt{2}}
\not\!{\epsilon^*}(P, S_z) (\not\!{P} + M)\ ,
\label{proj1}
\end{eqnarray}
respectively, for the  pseudoscalar and the vector quarkonium states. 
Here $\epsilon^{\mu}(P, S_z)$ denotes the polarization vector of 
the spin-1 quarkonium state, and $M = 2 m_c$ is the mass of it. Here,
$m_c$ is the charm quark mass. Projectors (\ref{proj0}) and (\ref{proj1})
map a $Q\bar{Q}$ pair into the S-wave states. In our study we also repeat 
previous calculations, where the $\jp$ prompt production was considered; 
that is, the $\jp$ coming from higher excited states feeddown is taken 
into consideration. For $P$-wave states production, to leading order 
one needs to expand the relative momentum of heavy quarks to first order. 
Of the spin projectors they are
\begin{eqnarray}
{\cal P}_{0,0}^\alpha(P; 0) = \frac{1}{2\sqrt{2} M}[\gamma^\alpha
\gamma_5 (\not\!{P} + M) + \gamma_5 (\not\!{P} + M) \gamma^\alpha ]
\label{proj2}\; , \\
{\cal P}_{1,S_z}^\alpha (P; 0) = 
\frac{1}{2\sqrt{2} M}[\gamma^\alpha
\not\!{\epsilon^*} (\not\!{P} + M) + 
\not\!{\epsilon^*}(\not\!{P} + M) \gamma^\alpha ]\; ,
\label{proj3}
\end{eqnarray}
respectively. 

With the above spin projectors, amplitudes for  
$\gamma + \gamma \ra J/\psi + c\;+\; \bar{c}$ process can be
obtained, which are presented in the Appendix
for reference and comparison. Nevertheless, the matrix 
element squared is too lengthy to be shown here. All the calculation 
is evaluated by using the automatic processing code for computers, 
the FDC \cite{wang:fdc96}. Interested readers, who want to have the 
lengthy expressions and the corresponding Fortran program,
are encouraged either to download directly from the web site, 
or write us.

\section{Numerical Results}

As stated in above, we perform the calculation of Feynman
diagram algebra by computerized program, in which  
the spin projectors method was built in and is more suitable for evaluating
complicated processes. The Feynman diagram, Analytic formulas 
and it's Fortran source 
are generated by the FDC. This program was employed in past 
in calculating the $\jp$ electromagnetic production at 
electron-positron colliders \cite{chang:yr97}, and in many 
other applications. In order to further secure the applicability 
of this program, in preparing this work we repeat several other 
independent processes and compared with the results given in literature. 
The numerical calculation is performed in batch by the Monta Carlo 
subroutine encoded in FDC, too.

The overall differential cross section of the $\jp$ photoproduction 
can be obtained by the double convolution of parton-parton 
(photon-photon) to $\jp$ cross section with the parton distribution 
functions in photon and the photon distribution densities, 
schematically like
\begin{eqnarray}
{d\sigma} = \int  dx_1\ dx_2\ dt\ f_\gamma(x_1)\ 
f_\gamma(x_2) \sum_{i,j, k} \int dx_i dx_j f_{i/\gamma}(x_i)\ 
f_{j/\gamma}(x_j)\ d\sigma_{i+ j \ra \psi + k} (x_i, x_j)\ 
\end{eqnarray}
Here, the $f_\gamma(x)$ represents the photon density in
$e^+\ e^-$ collisions or at photon colliders, and $f_{i/\gamma}(x)$
$(i,j=\gamma,g,u,d,s)$
denotes the GRS photon PDFs \cite{gluck:99yr}. 
For direct photon-photon
interaction, it is obvious that the $f_{\gamma/\gamma}(x)$ will be
the Delta function.

In doing numerical calculation, the general parameters are 
taken as:
$\alpha = 1/137.065$;\
$<{\cal O}^{J/\psi}(^3S_1^{[1]})> = 1.4\; {\rm GeV}^3$;\
$m_c = 1.5\pm 0.1$ GeV;\ 
$\Lambda_{QCD}^{(4)} = 174$ MeV \cite{Martin:yr98}  
and the strong coupling
is running with renormalization scale, $\mu= m_T$. Here,
$m_T = \sqrt{m_\psi^2 + p_T^2}$ is the normally defined 
transverse mass of $\jp$. 
The factor ${(J/\psi+\psi^{\prime})}=1.278$ is used to include  
$\psi^{\prime}$ contribution. 
With taking the non-relativistic
limit the $J/\psi$ mass is taken to be two times of charm
quark mass, and the open charm pair is of the same mass as charm quark in $J/\psi$, 
otherwise the gauge invariance will be broken.  
To keep the consistency with other
analyses, our choice of parameters is in accordance with
what taken in ref. \cite{klasen:yr02}. For details of 
the choice making, e.g, 
the magnitudes of color-octet matrix
elements, readers are recommended to
refer to the CTEQ5L fit used by \cite{klasen:yr02}.

\begin{figure}[tbh]
\begin{center}
\epsfig{file=plot1.ps,
bbllx=155pt,bblly=60pt,bburx=400pt, 
bbury=510pt,width=6cm,height=7.5cm,clip=0}
\end{center}
\caption[bt]{The transverse momentum distribution of $\jp$ 
photoproduction in LEP II experiment. The results of 
$\gamma + \gamma \rightarrow \jp + c+ \bar{c}$ process are 
confronted with the central values in previous study in Ref. 
\cite{klasen:yr02} and recent DELPHI experimental result 
\cite{delphi:yr01}.}
The upper bound of the shaded band is obtained at   
renormalization scale r=0.5($M_T$) and $m_c$=1.4,
and the lower one at r=2 and $m_c$=1.6 
\label{graph2}
\end{figure}    

In figure \ref{graph2} we present our process versus the
other theoretical predictions \cite{klasen:yr02} and recent
DELPHI experimental result \cite{delphi:yr01}, where in 
avoiding the non-photoproduction process the invariant mass 
of $\gamma\ \gamma$ system is limited to $W \le 35$ GeV as
performed in experiment. The maximum angle ensuring the
photon to be real in eq.(2), the $\theta_{\rm max}$,
is taken to be 32 mrad. As shown in the figure, the 
previously considered leading order CS processes contribute 
less than the process we are considering here as transverse 
momentum being larger than 1 GeV. When transverse momentum is 
small we know the diffractive interaction process would take 
the leading contribution for $\jp$ production.
>From the figure, the  discrepancy between experimental 
data and colour-singlet calculations is 
shrunk after including the 
$\gamma + \gamma \rightarrow \jp + c+ \bar{c}$ process,
although still the CS contributions fall below the data 
even with the optimal choice of the errors. In drawing
figure \ref{graph2}, FDC was used to re-calculated the former NRQCD and CSM 
processes done in \cite{klasen:yr02} and got an agreement with 
their results numerically.

Appearing in the figure, the results from previous study
are only taken by the central values. We notice that there 
are big uncertainties remaining in both previous analyses 
and our calculation shown as shaded band. The theoretical
errors come mainly from the influence of scale dependence 
($\mu=0.5m_T,m_T,2m_T$), the non-perturbative matrix element 
uncertainty, and the variation of charm quark mass 
($m_c=1.5\pm0.1GeV$). The strong scale and mass 
dependence imply that the higher order relativistic and 
radiative corrections would be large. 

As shown in the figure 2, the 
$\gamma + \gamma \rightarrow \jp + c+ \bar{c}$
process can not be reproduced by the fragmentation
mechanism  \cite{fragmentation:yr93} (simply times $2.4\times10^{-4}$ 
to $\gamma+\gamma \rightarrow c + \bar{c}$), even at the high $p_T$ side.
This finding means that the non-fragmentation-like graphs
existing in this process are not negligible in high
energy and high $p_T$ limit. This character hints that 
the $B_c$ photoproduction, hadroproduction
as well, with the similar topology of Feynman graphs can
not be simply replaced by fragmentation mechanism.

In figure 3 we present the invariant mass, angular, rapidity
and pseudo-rapidity distributions of 
$\gamma + \gamma \rightarrow \jp + c+ \bar{c}$ 
process, respectively. The invariant mass of colliding
photons starts at 6 GeV as physical requirement
and ends up at 35 GeV as imposed in DELPHI experiment.
The (pseudo)rapidity varies from -2 to 2 as performed
in experiment as well.

\begin{figure}[tbh]
\begin{center}
\epsfig{file=plot2.ps,
bbllx=155pt,bblly=60pt,bburx=400pt, 
bbury=510pt,width=6cm,height=7.5cm,clip=0}
\end{center}
\caption[bt]{From the upper left to lower right, it is the
plot of mass distribution, $d\sigma/dM_{\gamma\gamma}$; angular 
distribution, $d\sigma/d\cos\theta$; rapidity distribution, 
$d\sigma/d y$; pseudo-rapidity distribution, $d\sigma/d\eta$.}
\label{graph3}
\end{figure}    

To show the influence of the renormalization scale and charm 
quark mass, in Table I we present the relation of total cross 
section relying on them. Here, the $r$ means the fraction of 
renormalization scale $\mu$ on charmonium transverse mass. That 
is: $\mu = r\; m_T$, where $p_T$ is the $\jp$ transverse momentum. 
The total cross sections are obtained under the prerequisite of: 
$\sqrt(10)>p_T > 1.0$ GeV, $W < 35$ GeV, and $|\eta| <2$, where $W$ and 
$\eta$ are final states invariant mass and $\jp$ psudo-rapidity, 
respectively. 
  
\vskip 8mm
{\small{TABLE I. Renormalization scale and charm quark mass 
dependence of the total cross-section.}}
\begin{center}
\begin{tabular}{cccc}\hline\hline
\rule[-1.2ex]{0mm}{4ex}&\hspace{-1.4cm} $m_c = 1.4$ GeV 
\hspace{1cm}&
$m_c = 1.5$ GeV \hspace{1cm} & $m_c = 1.6$ GeV \\ \hline
\rule[-1.2ex]{0mm}{4ex} r = 0.5 \hspace{1cm} & $0.82 \; \mbox{pb}$ 
\hspace{1cm}
& $0.54\; \mbox{pb}$ \hspace{1cm} & $0.37\; \mbox{pb}$\\ \hline
\rule[-1.2ex]{0mm}{4ex} r = 1 \hspace{1cm} & $0.47\; \mbox{pb}$ 
\hspace{1cm} 
& $0.31\; \mbox{pb}$ \hspace{1cm} & $0.21\; \mbox{pb}$\\ \hline
\rule[-1.2ex]{0mm}{4ex} r = 2 \hspace{1cm} & $0.30 \; \mbox{pb}$ 
\hspace{1cm}
& $0.20\; \mbox{pb}$ \hspace{1cm} & $0.14\; \mbox{pb}$\\ 
\hline\hline
\end{tabular}
\end{center}
\vskip 5mm
It is evident that both renormalization scale and charm quark 
mass induce large uncertainties on total cross section. Among 
them, the scale dependence is basically more important. Under 
the same physical cut and parameter input we find the new process 
at LC(s), e.g. TESLA, gives a total cross section of 0.41 pb at 
the central values of scale and charm quark mass, which is larger 
than the total cross section at LEP as shown in Table I. This means
that the new process can surely be measured with high accuracy at 
future LCs.

\section{Discussion and Conclusions}

In this work we proceed the calculation of 
$\gamma + \gamma \ra J/\psi + c\;+\; \bar{c}$ process of
the $\jp$ photoproduction in $e^+\ e^-$ interactions in 
LEP II experiment, which was not taken into consideration 
in previous analyses. We find this process is quite
large in the photon-photon collision at LEP. The importance 
of including this process lies in two aspects:
1) it is the dominant precess among all the CS subprocesses
with transverse momentum greater than 1 GeV. Since in the low
$p_T$ region the diffractive interaction will be overwhelmingly
large, to have a clear cut of it one needs to focus on the large
$p_T$ area; 2) considering the large uncertainties still remain
in the Color-Octet matrix elements, in goal of either to fix these 
uncertainties, or to test the accuracy of QCD perturbative 
calculations, including this new process is very necessary.

The new process we considered is unique relative to other 
processes of $\jp$ production in $e^+\; e^-$ scattering. 
It should be obvious to obtain gauge invariance for the total amplitude.
We performed the check and found there is gauge invariance.
It is also checked for other processes 
in the aim of convincing us more of our results.
In practice, we performed the 
gauge invariance check in two different ways. One at the
amplitude level, we replace the photon polarization vector(s) 
by the corresponding momentum, then reallocate the independent
terms and then numerically calculate the amplitude square;
in another way, we do the replacement at matrix element square
level and then evaluate it. In both cases the large cancellation
happens, and up to the precision limit of Fortran program they
get to be zero.

We found that the uncertainty induced by 
scale variation is quite large, which means the higher order 
corrections could be big. The uncertainties remaining in the
quarkonium non-perturbative matrix elements and charm quark
mass are also the sources of theoretical predication errors.

We also compared the pure fragmentation 
result with our full calculation, and find that the 
situation of quarkonium production here differs from what 
in $Z^0$ decays, where the fragmentation scheme can almost 
reproduce the full Feynman diagram calculation results. It 
is noticed that after considering our proposed process, the 
preliminary DELPHI data are still not explainable by CSM 
alone, and colour-octet scenario is still necessary. 
Nevertheless, since large uncertainties remains
in both CS and NRQCD analyses, quantitative conclusions for
the universality of colour-octet matrix elements still hard to
get. In our opinion, to have a full NLO calculation of prompt
$\jp$ production would be critical on this point and beyond. 
Although the present DELPHI data are just marginal in observing
the $\gamma + \gamma \ra J/\psi + c + \bar{c}$ signal, it would 
be still interesting for experimenters to see whether the 
accumulated LEP data are enough or not to find it. Anyhow,
at LCs this new process should be observed with a high
precision.
It is also noticed that in previous NLO calculation 
of $\jp$ photoproduction at HERA \cite{kramer:yr96,kramer:95}, 
the similar same order process 
$g + \gamma \ra J/\psi(\psi') + c + \bar{c}$  
was missing. Whereas, naive estimation tells us the missing 
part should be not so important as the case in photon-photon 
interactions. Detailed investigation on this will be presented 
elsewhere.

\vspace{0.5cm}
\centerline{\bf Acknowledgments}
C.-F.Q. would like to thank J.P. Ma for discussions over the 
related issues and thank ITP for their hospitality while part 
of this work was done. This work was supported in part by the 
National Science Foundation of China with Contract No.19805015 
and No.90103013.

\newpage 
\centerline{\bf Appendix}
\setcounter{equation}{0}
\renewcommand{\theequation}{A-\arabic{equation}}
The matrix element of process $\gamma + \gamma \rightarrow \jp + c+ \bar{c}$. 
Here, $\epsilon_{1},\ \epsilon_{2}\ \epsilon_{3}$ are the polarizations of 
initial photons and $\jp$, respectively; and $p_i(i = 1, 2, 3)$ are the 
corresponding momenta of them.
\begin{equation}\begin{array}{ll}
 M&=c~\bar{u}(p_4)(c_{1}{\hat {p_3}}{\hat {\epsilon_{3}}}{\hat {p_1}}
{\hat {\epsilon_{1}}}+
c_{2}{\hat {p_3}}{\hat {\epsilon_{3}}}+c_{3}{\hat {\epsilon_{2}}}{\hat {p_1}}
{\hat {p_3}}{\hat {\epsilon_{1}}}+c_{4}{\hat {\epsilon_{2}}}{\hat {p_1}}{\hat 
{\epsilon_{3}}}{\hat {\epsilon_{1}}}+c_{5}{\hat {\epsilon_{2}}}{\hat {p_3}}
{\hat {\epsilon_{3}}}{\hat {p_1}}{\hat {\epsilon_{1}}}+c_{6}{\hat 
{\epsilon_{2}}}{\hat {p_3}}{\hat {\epsilon_{3}}}{\hat {p_1}} \\ &+c_{7}{\hat 
{\epsilon_{2}}}{\hat {p_3}}{\hat {\epsilon_{3}}}{\hat {\epsilon_{1}}}+c_{8}
{\hat {\epsilon_{2}}}{\hat {\epsilon_{1}}}+c_{10}{\hat {\epsilon_{2}}}{\hat 
{p_3}}{\hat {\epsilon_{1}}}+c_{11}{\hat {\epsilon_{2}}}{\hat {p_3}}{\hat 
{\epsilon_{3}}}+c_{12}{\hat {\epsilon_{2}}}{\hat {p_3}} +c_{13}{\hat 
{\epsilon_{2}}}{\hat {\epsilon_{3}}}{\hat {\epsilon_{1}}}+c_{14}{\hat 
{\epsilon_{2}}}{\hat {\epsilon_{3}}}+ c_{15}{\hat {\epsilon_{2}}}{\hat {p_1}}
{\hat {\epsilon_{1}}}\\&+c_{16}{\hat {\epsilon_{2}}}{\hat {p_1}}{\hat 
{\epsilon_{3}}}+c_{17}{\hat {\epsilon_{2}}}+c_{18}{\hat {\epsilon_{3}}}{\hat 
{p_1}}{\hat {\epsilon_{1}}}+c_{19}{\hat {\epsilon_{3}}}{\hat {\epsilon_{1}}}+c_{
20}{\hat {\epsilon_{3}}}+c_{21} +c_{22}{\hat {p_1}}{\hat {\epsilon_{1}}}+
c_{23}{\hat {p_3}}{\hat {p_1}}{\hat {\epsilon_{1}}}+c_{24}{\hat {p_3}}{\hat 
{\epsilon_{1}}}\\&+c_{25}{\hat {p_3}}+c_{26}{\hat {\epsilon_{1}}}+c_{27}{\hat {p_1}
}+c_{28}{\hat {p_3}}{\hat {p_1}}+c_{29}{\hat {\epsilon_{3}}}{\hat {p_1}}+c_{30}
{\hat {p_1}}{\hat {p_3}}{\hat {\epsilon_{3}}} +c_{31}{\hat {\epsilon_{2}}
}{\hat {p_1}}{\hat {p_3}}+c_{32}{\hat {\epsilon_{2}}}{\hat {p_1}}
)v(p_5)
\end{array}\end{equation}
where 
\begin{equation}\begin{array}{llll}
&c^2={\displaystyle{4096\alpha^2{\alpha_s}^2
<{\cal O}^{J/\psi}(^3S_1^{[1]})> 
{(J/\psi+\psi^{\prime})}
{\pi}^4 \over 6561{m_{c}}}}
& c_i=\sum_{j=1,20}~c_{i,j}~y_j~&
\end{array}\end{equation}
\begin{equation}\begin{array}{lll}
&y_{1}=(4x_{3}x_{10}(2{m^2_{c}}-x_{8}-2x_{10}+x_{15}))^{-1}
&y_{2}=(-2x_{3}(2{m^2_{c}}+x_{15})(4{m^2_{c}}+2x_{15}))^{-1}\\
&y_{3}=(2x_{3}x_{8}(2{m^2_{c}}+x_{15}))^{-1}
&y_{4}=(2x_{3}x_{8}(2{m^2_{c}}-x_{2}-2x_{3}+x_{14}))^{-1}\\
&y_{5}=(-2x_{4}(2{m^2_{c}}+x_{14})(4{m^2_{c}}+2x_{14}))^{-1}
&y_{6}=(2x_{4}x_{8}(2{m^2_{c}}+x_{14}))^{-1}\\
&y_{7}=(2x_{4}x_{8}(2{m^2_{c}}-x_{2}-2x_{4}+x_{15}))^{-1}
&y_{8}=(-x_{8}(2{m^2_{c}}+x_{15})(2{m^2_{c}}+x_{14}+x_{15}+2x_{19}))^{-1}\\
&y_{9}=(-x_{8}(2{m^2_{c}}+x_{14})(2x_{1}-x_{2}-x_{8}))^{-1}
&y_{10}=(-x_{2}(2{m^2_{c}}+x_{14})(2x_{1}-x_{2}-x_{8}))^{-1}\\
\end{array}\end{equation}
\begin{equation}\begin{array}{ll}
&x_{1}={p_1}\cdot {p_2},~
x_{2}={p_1}\cdot {p_3},~
x_{3}={p_1}\cdot {p_4},~
x_{4}={p_1}\cdot {p_5},~
x_{5}={p_1}\cdot {\epsilon_{1}},~
x_{6}={p_1}\cdot {\epsilon_{2}},~\\&
x_{7}={p_1}\cdot {\epsilon_{3}},~
x_{8}={p_2}\cdot {p_3},~
x_{9}={p_2}\cdot {p_4},~
x_{10}={p_2}\cdot {p_5},~
x_{11}={p_2}\cdot {\epsilon_{1}},~
x_{12}={p_2}\cdot {\epsilon_{2}},~\\&
x_{13}={p_2}\cdot {\epsilon_{3}},~
x_{14}={p_3}\cdot {p_4},~
x_{15}={p_3}\cdot {p_5},~
x_{16}={p_3}\cdot {\epsilon_{1}},~
x_{17}={p_3}\cdot {\epsilon_{2}},~
x_{18}={p_3}\cdot {\epsilon_{3}},~\\&
x_{19}={p_4}\cdot {p_5},~
x_{20}={p_4}\cdot {\epsilon_{1}},~
x_{21}={p_4}\cdot {\epsilon_{2}},~
x_{22}={p_4}\cdot {\epsilon_{3}},~
x_{23}={p_5}\cdot {\epsilon_{1}},~
x_{24}={p_5}\cdot {\epsilon_{2}},~\\&
x_{25}={p_5}\cdot {\epsilon_{3}},~
x_{26}={\epsilon_{1}}\cdot {\epsilon_{2}},~
x_{27}={\epsilon_{1}}\cdot {\epsilon_{3}},~
x_{28}={\epsilon_{2}}\cdot {\epsilon_{3}},~
\end{array}\end{equation}
\begin{equation}\begin{array}{ll}
c_{1}=&(4x_{24}(4y_{10}+y_{3}+3y_{4}+3y_{6}+y_{7}+3y_{8}+4y_{9})+4x_{21}(-y_{3}+
y_{4}+y_{6}\\ &-y_{7}-y_{8})+4x_{17}(2y_{10}+2y_{4}+2y_{6}+y_{8}+2y_{9}))
\end{array}\end{equation}
\begin{equation}\begin{array}{ll}
c_{2}=&(x_{26}(4x_{8}(-y_{8}-y_{9})+8x_{4}(y_{10}+y_{6}+y_{7}+2y_{9})+8x_{3}(-
y_{6}-y_{7}-y_{9})\\ &+4x_{2}(y_{10}-2y_{6}-2y_{7})+8x_{1}(-y_{10}+y_{6}+
y_{7})+8(2{m_{c}}^2y_{8}-x_{10}y_{8}\\ &+x_{15}y_{8}))+(8x_{24}x_{23}(3y_{6}
+y_{7}+y_{8})+8x_{23}x_{21}(y_{6}-y_{7}+y_{8})+8x_{24}x_{20}( \\ 
&-y_{3}-3y_{4}+y_{8})+8x_{21}x_{20}(y_{3}-y_{4}+y_{8})+4x_{24}x_{16}
(-y_{10}+3y_{8}-y_{9}) \\ 
&+4 x_{21}x_{16}(y_{10}+y_{8}+y_{9})+16(-x_{17}x_{20}y_{4}+x_{17}x_{23}y_{6})))
\end{array}\end{equation}
\begin{equation}\begin{array}{ll}
c_{3}=&(4x_{7}(y_{10}-3y_{3}-y_{4}+y_{6}-y_{7}+y_{9})+4x_{25}(-2y_{10}+y_{3}-
y_{4}-y_{6}+y_{7}-3y_{8}-2y_{9})   \\  &+4x_{22}(2y_{10}+3y_{3}+y_{4}+y_{6}+3
y_{7}+y_{8}+2y_{9})+4x_{13}(y_{10}-2y_{3}-2y_{7}+y_{9}))
\end{array}\end{equation}
\begin{equation}\begin{array}{ll}
c_{4}=&(16{m_{c}}^2(-2y_{3}-2y_{7}+y_{8})+(4x_{8}(-y_{10}+2y_{3}+2y_{7}-y_{9})+4
x_{2}(-y_{10}\\ &+3y_{3}+y_{4}-y_{6}+y_{7}-y_{9})+4x_{15}(2y_{10}-y_{3}+
y_{4}+y_{6}-y_{7}+3y_{8}+2y_{9})\\ &+4x_{14}(-2y_{10}-3y_{3}-y_{4}-y_{6}-3
y_{7}-y_{8}-2y_{9})))
\end{array}\end{equation}
\begin{equation}\begin{array}{ll}
c_{5}=&4{m_{c}}(-y_{1}-y_{10}-y_{2}-y_{3}-y_{4}-y_{5}-y_{6}-y_{7}-y_{8}-y_{9})
\end{array}\end{equation}
\begin{equation}\begin{array}{ll}
c_{6}=&(4x_{23}(y_{10}-4y_{6}-4y_{7}+3y_{8}-y_{9})+4x_{20}(-3y_{10}-4y_{6}-4
y_{7}\\ &-y_{8}-5y_{9})+4x_{16}(-y_{10}-4y_{6}-4y_{7}+y_{8}-3y_{9}))
\end{array}\end{equation}
\begin{equation}\begin{array}{ll}
c_{7}=&(4x_{8}(y_{8}+y_{9})+8x_{4}(-y_{10}+y_{6}+y_{7}+y_{8})+4x_{3}(2y_{10}+2
y_{3}+2y_{4} \\ &+4y_{6}+4y_{7}+y_{8} +4y_{9})+4x_{2}(4y_{6}+4y_{7}+3y_{9})+4
x_{1}(y_{10}-4y_{6}-4y_{7} \\ &-3y_{9})+4(-7{m_{c}}^2y_{8}+5x_{10}y_{8}-5
x_{15}y_{8}-3x_{19}y_{8}))
\end{array}\end{equation}
\begin{equation}\begin{array}{ll}
c_{8}=&({m_{c}}^2(16x_{7}(y_{10}+4y_{3}-y_{8}+y_{9})+96x_{25}y_{8})+(8x_{25}
x_{2}(-y_{3}+y_{4}-y_{8})   \\  &+8x_{22}x_{2}(-y_{10}-3y_{3}-y_{4}-y_{8}-y_{9})
+8x_{7}x_{15}(y_{3}-y_{4}+y_{8})+8x_{7}x_{14}(\\ &y_{10}+3y_{3}+y_{4}+y_{8}+
y_{9})+8(3x_{13}x_{15}y_{8}+2x_{13}x_{2}y_{3}+3x_{14}x_{25}y_{8}\\ 
&-3x_{15}x_{22} y_{8}-3x_{25}x_{8}y_{8}-2x_{7}x_{8}y_{3})))
\end{array}\end{equation}
\begin{equation}\begin{array}{ll}
c_{10}=&8{m_{c}}x_{7}(y_{10}+y_{5}+y_{6}+y_{7}+y_{9})
\end{array}\end{equation}
\begin{equation}\begin{array}{ll}
c_{11}=&{m_{c}}(8x_{23}(-y_{5}-y_{6}-y_{7}+y_{8})+8x_{20}(y_{1}+y_{2}+y_{3}+
y_{4}+y_{8})+4x_{16}(y_{10}+y_{8}+y_{9}))
\end{array}\end{equation}
\begin{equation}\begin{array}{ll}
c_{12}=&(x_{27}(4x_{8}(-y_{8}-y_{9})+8x_{4}(2y_{10}-y_{6}-y_{7})+8x_{3}(-y_{10}-
3y_{6}-3y_{7}-3y_{9})   \\  &+4x_{2}(y_{10}-6y_{6}-6y_{7}-4y_{9})+8x_{1}(-y_{10}
+3y_{6}+3y_{7}+2y_{9})+8(2{m_{c}}^2y_{8}   \\  &-x_{10}y_{8}+x_{15}y_{8}))+(8
x_{7}x_{23}(-2y_{10}+2y_{6}+4y_{7}-3y_{8})+8x_{25}x_{23}(y_{6}\\ &-y_{7} +
y_{8})+8x_{23}x_{22}(-y_{6}-3y_{7}+3y_{8})+8x_{7}x_{20}(y_{10}-3y_{3}-y_{4} \\ 
&+3y_{6}+3y_{7} +3y_{9})+8x_{25}x_{20}(y_{3}-y_{4}-y_{8})+8x_{22}x_{20}(3
y_{3}+y_{4}+y_{8}) \\ &+4x_{7}x_{16}(-y_{10}+6y_{6}+6y_{7}-y_{8}+3y_{9})+4
x_{25}x_{16}(-y_{10}-y_{8}-y_{9}) \\ &+4x_{22}x_{16}(y_{10}+y_{8}+y_{9})+16
x_{23}x_{13}(y_{7}-y_{8})-16x_{13}x_{20}y_{3}))
\end{array}\end{equation}
\begin{equation}\begin{array}{ll}
c_{13}=&({m_{c}}(12x_{8}(-y_{8}-y_{9})+8x_{4}(y_{10}-2y_{6}-2y_{7}-y_{8}-y_{9})+
8x_{3}(y_{1}-2y_{10}-y_{3}-y_{4}   \\  &-3y_{6}-3y_{7}-4y_{9})+4x_{2}(-3y_{10}-2
y_{5}-8y_{6}-8y_{7}-2y_{8}-10y_{9})+8x_{15}(4y_{8}+y_{9})   \\  &+24x_{1}(y_{6}+
y_{7}+y_{9})+8(-3x_{10}y_{8}+x_{14}y_{9}+x_{19}y_{8}+x_{9}y_{8}))+8{m_{c}}^3(7
y_{8}+4y_{9}))
\end{array}\end{equation}
\begin{equation}\begin{array}{ll}
c_{14}=&({m_{c}}^2(64x_{23}(y_{7}-y_{8})+16(-x_{16}y_{8}-4x_{20}y_{3}))+(16x_{8}
x_{23}(-y_{7}+y_{8})   \\  &+8x_{23}x_{2}(y_{10}-3y_{6}-5y_{7}+3y_{8}-y_{9})+8
x_{20}x_{2}(-2y_{10}+3y_{3}+y_{4}-4y_{6}-4y_{7}   \\  &-4y_{9})+4x_{8}x_{16}(
y_{8}+y_{9})+8x_{4}x_{16}(-2y_{10}+y_{6}+y_{7})+8x_{3}x_{16}(y_{10}+3y_{6}
   \\  &+3y_{7}+3y_{9})+4x_{2}x_{16}(-2y_{10}-2y_{6}-2y_{7}+y_{8}-y_{9})+8x_{23}
x_{15}(-y_{6}+y_{7})   \\  &+8x_{20}x_{15}(-y_{3}+y_{4}+2y_{8})+4x_{16}x_{15}(
y_{10}+y_{8}+y_{9})+8x_{23}x_{14}(y_{6}+3y_{7}   \\  &-3y_{8})+8x_{20}x_{14}(-3
y_{3}-y_{4}-y_{8})+4x_{16}x_{14}(-y_{10}-y_{8}-y_{9})+8x_{16}x_{1}(y_{10}
   \\  &-3y_{6}-3y_{7}-2y_{9})+8(x_{10}x_{16}y_{8}+2x_{20}x_{8}y_{3})))
\end{array}\end{equation}
\begin{equation}\begin{array}{ll}
c_{15}=&{m_{c}}(8x_{7}(y_{1}-2y_{3}+y_{6}-y_{7})+8x_{25}(-y_{10}+y_{2}+y_{3}-
y_{4}-y_{6}+y_{7}-2y_{8}\\ &-y_{9}) +8x_{22}(-y_{1}+2y_{10}+2y_{3}-y_{5}+2
y_{7}+y_{8}+2y_{9})+8x_{13}(-y_{3}-y_{7}))
\end{array}\end{equation}
\begin{equation}\begin{array}{ll}
c_{16}=&{m_{c}}(8x_{23}(y_{10}-3y_{6}-3y_{7}+2y_{8}-y_{9})+8x_{20}(-2y_{10}-3
y_{6}-3y_{7}\\ &-y_{8}-4y_{9})+4x_{16}(-3y_{10}-2y_{5}-8y_{6}-8y_{7}+y_{8}-7
y_{9}))
\end{array}\end{equation}
\begin{equation}\begin{array}{ll}
c_{17}=&(x_{27}({m_{c}}(8x_{8}(y_{8}+2y_{9})+16x_{4}(-2y_{10}+y_{6}+y_{7})+32
x_{3}(y_{6}+y_{7}+y_{9})   \\  &+16x_{2}(-y_{10}+2y_{6}+2y_{7}+2y_{9})+16x_{15}(
-y_{8}-y_{9})+16x_{1}(y_{10}-2y_{6}-2y_{7}\\ &-y_{9})+16(x_{10}y_{8}-x_{14}
y_{9}))+32{m_{c}}^3(-y_{8}-2y_{9}))+{m_{c}}(16x_{7}x_{23}(y_{6}-y_{7}+y_{8})
   \\  &+16x_{25}x_{23}(-y_{6}+y_{7}-y_{8})+16x_{23}x_{22}(-y_{5}+2y_{7}-3y_{8})
+16x_{7}x_{20}(-y_{1}   \\  &+y_{10}+2y_{3}+y_{9})+16x_{25}x_{20}(-y_{2}-y_{3}+
y_{4}+2y_{8})+16x_{22}x_{20}(y_{1}-2y_{3})   \\  &+8x_{7}x_{16}(y_{10}+y_{8}+
y_{9})+8x_{25}x_{16}(y_{10}+y_{8}+y_{9})+16x_{23}x_{13}(-y_{7}+2y_{8})   \\  &+
16x_{20}x_{13}(y_{3}-y_{8})+8(x_{13}x_{16}y_{8}-2x_{16}x_{22}y_{8})))
\end{array}\end{equation}
\begin{equation}\begin{array}{ll}
c_{18}=&{m_{c}}(8x_{24}(4y_{10}+y_{3}+3y_{4}+3y_{6}+y_{7}+2y_{8}+4y_{9})+8x_{21}
(y_{10}+2y_{4}+2y_{6}\\ &-y_{8}+y_{9})+4x_{17}(5y_{10}+2y_{2}+3y_{3}+5y_{4}+
5y_{6}+3y_{7}+3y_{8}+5y_{9}))
\end{array}\end{equation}
\begin{equation}\begin{array}{ll}
c_{19}=&(8{m_{c}}^2(5x_{17}y_{8}-4x_{24}y_{8})+(16x_{24}x_{2}(2y_{10}+2y_{6}+
y_{7}+y_{8}+2y_{9})   \\  &+8x_{21}x_{2}(y_{10}+2y_{6}+y_{9})+4x_{8}x_{17}(-
y_{8}-y_{9})+16x_{4}x_{17}(-y_{7}-y_{8})   \\  &+8x_{3}x_{17}(-y_{10}-y_{3}-
y_{4}-y_{6}-3y_{7}-y_{8}-y_{9})+4x_{2}x_{17}(4y_{10}+4y_{6}\\ 
&-4y_{7}+y_{8}
+3y_{9})+8x_{17}x_{1}(y_{6}+3y_{7}+y_{9})+8(-4x_{10}x_{17}y_{8}-x_{14}x_{24}
y_{8} \\ 
&+3x_{15}x_{17}y_{8}-x_{15}x_{21}y_{8}-2x_{15}x_{24}y_{8}+3x_{17}
x_{19}y_{8}+x_{24}x_{8}y_{8})))
\end{array}\end{equation}
\begin{equation}\begin{array}{ll}
c_{20}=&(x_{26}({m_{c}}(8x_{8}(-y_{8}-2y_{9})+16x_{4}(y_{10}+y_{6}+y_{7}+2y_{9})
+16x_{15}(y_{8}+y_{9})   \\  &+16x_{1}(-y_{10}-y_{9})+8(-2x_{10}y_{8}+2x_{14}
y_{9}+x_{2}y_{10}))+32{m_{c}}^3(y_{8}+2y_{9}))   \\  &+{m_{c}}(16x_{24}x_{23}(3
y_{6}+y_{7}-y_{8})+16x_{24}x_{20}(-y_{3}-3y_{4})+16x_{21}x_{20}(-2y_{4}   \\  &+
y_{8})+8x_{23}x_{17}(5y_{6}+3y_{7}-3y_{8}-2y_{9})+8x_{20}x_{17}(-2y_{2}-3y_{3}-5
y_{4}\\ 
&-y_{8}-2y_{9})+8x_{24}x_{16}(-3y_{10}-y_{8}-3y_{9})+16x_{21}x_{16}(
-y_{10}-y_{9}) \\ 
&+4x_{17}x_{16}(-5y_{10}-3y_{8}-9y_{9})+32x_{21}x_{23}y_{6}))
\end{array}\end{equation}
\begin{equation}\begin{array}{ll}
c_{21}=&(x_{26}(32{m_{c}}^2x_{7}(y_{10}-2y_{6}+2y_{7}-2y_{9})+(16x_{8}x_{7}(
y_{6}-y_{7}+y_{9})+32x_{25}x_{2}(   \\  &y_{6}+y_{9})+16x_{22}x_{2}(-y_{10}-2
y_{7})+32x_{7}x_{15}(-y_{6}-y_{9})+16x_{7}x_{14}(y_{10}+2y_{7})   \\  &+16x_{2}
x_{13}(-y_{6}+y_{7}-y_{9})))+x_{27}({m_{c}}^2(32x_{17}(-y_{8}+y_{9})+64x_{24}
y_{8})   \\  &+(16x_{24}x_{2}(-3y_{10}-4y_{6}-2y_{7}-y_{8}-3y_{9})+16x_{21}x_{2}
(-y_{10}-2y_{6}-y_{9})   \\  &+16x_{3}x_{17}(y_{10}+y_{6}+y_{7}+y_{9})+8x_{2}
x_{17}(-3y_{10}-4y_{6}-3y_{9})+8x_{17}x_{15}(-2y_{8}   \\  &+y_{9})+16x_{17}
x_{1}(-y_{6}-y_{7}-y_{9})+8(2x_{10}x_{17}y_{8}+x_{14}x_{17}y_{9}+2x_{14}x_{24}
y_{8}   \\  &+2x_{15}x_{24}y_{8}+x_{17}x_{8}y_{8}-2x_{24}x_{8}y_{8})))+x_{28}(
{m_{c}}^2(32x_{23}(y_{6}-y_{7}+y_{8})   \\  &+32x_{20}(y_{3}-y_{4}+y_{8})+16
x_{16}(y_{10}+3y_{8}-y_{9}))+(16x_{23}x_{2}(y_{10}+3y_{7}-y_{8}   \\  &+y_{9})+
16x_{20}x_{2}(2y_{10}-y_{3}+y_{6}+3y_{7}+2y_{9})+16x_{4}x_{16}(y_{10}+y_{6}+
y_{7}+y_{9})   \\  &+8x_{2}x_{16}(4y_{10}+2y_{6}+6y_{7}-y_{8}+4y_{9})+8x_{16}
x_{15}(2y_{8}-y_{9})+16x_{23}x_{14}(-y_{7}   \\  &+y_{8})+16x_{20}x_{14}(y_{3}+
y_{8})+8x_{16}x_{14}(y_{10}+y_{8})+8(-2x_{1}x_{16}y_{10}   \\  &-2x_{10}x_{16}
y_{8}-2x_{15}x_{20}y_{4}+2x_{15}x_{23}y_{6}-x_{16}x_{8}y_{8}+2x_{20}x_{8}y_{4}-2
x_{23}x_{8}y_{6})))   \\  &+(16x_{7}x_{23}x_{17}(-3y_{7}+y_{8})+16x_{23}x_{22}
x_{17}(y_{7}-y_{8})+16x_{7}x_{20}x_{17}(-y_{10}   \\  &+y_{3}-y_{6}-3y_{7}-y_{9}
)+16x_{22}x_{20}x_{17}(-y_{3}-y_{8})+16x_{7}x_{24}x_{16}(2y_{10}+3y_{6}+y_{7}
   \\  &+y_{8}+2y_{9})+16x_{7}x_{21}x_{16}(y_{6}-y_{7})+8x_{7}x_{17}x_{16}(
y_{10}+2y_{6}-6y_{7}+y_{8}+y_{9})   \\  &+8x_{22}x_{17}x_{16}(-y_{10}-y_{8}-
y_{9})+16(x_{13}x_{16}x_{24}y_{8}-x_{13}x_{17}x_{20}y_{4}   \\  &+x_{13}x_{17}
x_{23}y_{6}-x_{16}x_{22}x_{24}y_{8}-x_{16}x_{24}x_{25}y_{8}+x_{17}x_{20}x_{25}
y_{4}-x_{17}x_{23}x_{25}y_{6})))
\end{array}\end{equation}
\begin{equation}\begin{array}{ll}
c_{22}=&(x_{28}(16{m_{c}}^2(y_{10}-y_{3}+y_{4}+y_{6}-y_{7}+y_{8}+y_{9})+(8x_{8}(
-y_{10}-y_{4}-y_{6}-y_{9})   \\  &+8x_{2}(-y_{10}+y_{3}-y_{6}-y_{9})+8x_{15}(
y_{10}+y_{4}+y_{6}+2y_{8}+y_{9})+8x_{14}(-y_{3} \\ 
&-y_{7}-y_{8}))) +(8x_{7} x_{17}(y_{10}-y_{3}+y_{6}+y_{9})+8x_{25}x_{17}(
-y_{10}-y_{4}-y_{6} \\ 
&-2y_{8}-y_{9})+8x_{22}x_{17}(y_{3}+y_{7}+y_{8})+8x_{17}x_{13}(y_{10}
+y_{4}+y_{6}+y_{9})))
\end{array}\end{equation}
\begin{equation}\begin{array}{ll}
c_{23}=&8x_{28}{m_{c}}(-y_{2}-y_{3}-y_{7}-y_{8})
\end{array}\end{equation}
\begin{equation}\begin{array}{ll}
c_{24}=&(x_{28}(4x_{8}(y_{8}+y_{9})+16x_{4}(y_{7}+y_{8})+8x_{3}(y_{10}+y_{3}+
y_{4}+y_{6}+3y_{7}+y_{8}+y_{9})   \\  &+4x_{2}(y_{10}+2y_{6}+6y_{7}+2y_{9})+8
x_{1}(-y_{6}-3y_{7}-y_{9})+8(-5{m_{c}}^2y_{8}+4x_{10}y_{8}   \\  &-4x_{15}y_{8}-
3x_{19}y_{8}))+(16x_{7}x_{24}(-2y_{10}-2y_{6}-y_{7}-y_{8}-2y_{9})+8x_{7}x_{21}(
   \\  &-y_{10}-2y_{6}-y_{9})+4x_{7}x_{17}(-5y_{10}-6y_{6}-2y_{7}-y_{8}-5y_{9})+
8(-x_{13}x_{24}y_{8}   \\  &+x_{17}x_{25}y_{8}+x_{21}x_{25}y_{8}+x_{22}x_{24}
y_{8}+2x_{24}x_{25}y_{8})))
\end{array}\end{equation}
\begin{equation}\begin{array}{ll}
c_{25}=&{m_{c}}x_{28}(16x_{23}(-y_{7}+y_{8})+16x_{20}(y_{2}+y_{3}+y_{8})+8x_{16}
y_{8})
\end{array}\end{equation}
\begin{equation}\begin{array}{ll}
c_{26}=&(x_{28}({m_{c}}(8x_{8}(2y_{8}+y_{9})+32x_{4}(y_{7}+y_{8})+16x_{3}(y_{10}
+y_{4}+y_{6}+2y_{7}+y_{8}+y_{9})   \\  &+8x_{2}(y_{10}+2y_{6}+6y_{7}+2y_{8}+2
y_{9})+16x_{1}(-y_{6}-2y_{7}-y_{9})+16(3x_{10}y_{8}   \\  &-x_{14}y_{8}-4x_{15}
y_{8}-3x_{19}y_{8}))-112{m_{c}}^3y_{8})+{m_{c}}(16x_{7}x_{24}(-3y_{10}-3y_{6}-
y_{7}   \\  &-2y_{8}-3y_{9})+16x_{7}x_{21}(-y_{10}-2y_{6}-y_{9})+8x_{7}x_{17}(-4
y_{10}-5y_{6}-3y_{7}-y_{8}   \\  &-4y_{9})+16(-x_{13}x_{24}y_{8}+x_{17}x_{25}
y_{8}+x_{21}x_{25}y_{8}+x_{22}x_{24}y_{8}+2x_{24}x_{25}y_{8})))
\end{array}\end{equation}
\begin{equation}\begin{array}{ll}
c_{27}=&({m_{c}}x_{27}(16x_{24}(3y_{10}+3y_{6}+y_{7}+y_{8}+3y_{9})+16x_{21}(
y_{10}+2y_{6}+y_{9})   \\  &+8x_{17}(4y_{10}+5y_{6}+3y_{7}+y_{8}+4y_{9}))+
{m_{c}}x_{28}(16x_{23}(-y_{6}-2y_{7}+y_{8})   \\  &+16x_{20}(-y_{10}-y_{6}-2
y_{7}-y_{8}-y_{9})+8x_{16}(-y_{10}-2y_{6}-6y_{7}-y_{9}))   \\  &+{m_{c}}x_{26}(
16x_{7}(y_{6}-y_{7}+y_{9})+16x_{25}(-y_{6}+y_{7}-y_{8}-2y_{9})+16x_{22}( \\ 
&y_{10}- y_{5}+2y_{7})+16x_{13}(-y_{7}+y_{9})))
\end{array}\end{equation}
\begin{equation}\begin{array}{ll}
c_{28}=&(x_{27}(8x_{24}(3y_{10}+3y_{6}+y_{7}+y_{8}+3y_{9})+8x_{21}(y_{6}-y_{7})+
4x_{17}(3y_{10}+4y_{6}   \\  &+y_{8}+3y_{9}))+x_{28}(8x_{23}(-y_{10}-y_{6}-3
y_{7}+2y_{8}-y_{9})+8x_{20}(-2y_{10}-y_{6}-3y_{7}   \\  &-y_{8}-2y_{9})+4x_{16}(
-3y_{10}-2y_{6}-6y_{7}+y_{8}-3y_{9}))+x_{26}(8x_{7}(y_{6}-y_{7}+y_{9})   \\  &+8
x_{25}(-y_{6}+y_{7}-y_{8}-2y_{9})+8x_{22}(y_{10}+y_{6}+3y_{7}+y_{9})+8x_{13}(-2
y_{7}+y_{9})))
\end{array}\end{equation}
\begin{equation}\begin{array}{ll}
c_{29}=&(x_{26}(16{m_{c}}^2(-y_{10}+2y_{6}-2y_{7}+y_{8}+3y_{9})+(8x_{8}(-y_{6}+
y_{7}-2y_{9})+16x_{2}( \\ &-y_{6}-y_{9})+8x_{15}(2y_{6}+y_{8}+3y_{9})+8x_{14}
(-y_{10}-2y_{7})))+(8x_{23}x_{17}(y_{6}+3y_{7}   \\  &-2y_{8})+8x_{20}x_{17}(
y_{10}+y_{6}+3y_{7}+y_{8}+y_{9})+8x_{24}x_{16}(-3y_{10}-3y_{6}-y_{7} \\&-y_{8}
-3y_{9})+8x_{21}x_{16}(-y_{6}+y_{7})+8x_{17}x_{16}(-y_{10}-y_{6}+3y_{7}-
y_{8}-y_{9})))
\end{array}\end{equation}
\begin{equation}\begin{array}{ll}
c_{30}=&8x_{26}{m_{c}}(-y_{5}-y_{6}-y_{7}-y_{9})
\end{array}\end{equation}
\begin{equation}\begin{array}{ll}
c_{31}=&8x_{27}{m_{c}}(y_{10}+y_{5}+y_{6}+y_{7}+y_{9})
\end{array}\end{equation}
\begin{equation}\begin{array}{ll}
c_{32}=&(x_{27}(16{m_{c}}^2(y_{10}+4y_{7}-y_{8}+y_{9})+(8x_{2}(y_{6}-y_{7})+8
x_{15}(-y_{10}-y_{6}\\ &+y_{7}-y_{8} -y_{9})+8x_{14}(2y_{10}+y_{6}+3y_{7}+2
y_{9})-16x_{8}y_{7}))+(8x_{7}x_{16}(y_{10} \\ &+2y_{7}+y_{9}) +8x_{25}x_{16}(-
2y_{7}+y_{8})+8x_{22}x_{16}(-3y_{10}-2y_{6}-4y_{7}-3y_{9})\\ 
&+8x_{16}x_{13}(y_{10}+y_{6}+3y_{7}+y_{9})))
\end{array}\end{equation}
\end{document}